\input{aipcheck}

\documentclass[
    ,final            
  ]
  {aipproc}
\def \G{\mathcal{G}}
\def \T{\mathcal{T}}

\def \A{\mathcal{A}}

\usepackage{bm}     
\layoutstyle{8x11single}

\begin{document}

\title{Pairing in bulk nuclear matter beyond BCS}


\author{D. Ding}{
  address={Department of Physics, Washington University, 
St. Louis, Missouri 63130, USA}
}

\author{S. J. Witte}{
  address={Department of Physics, Washington University, 
St. Louis, Missouri 63130, USA},altaddress={Department of Physics and Astronomy,
University of California, Los Angeles, CA 90095, USA}
}

\author{W. H. Dickhoff}{
  address={Department of Physics, Washington University, 
St. Louis, Missouri 63130, USA}
}
\author{H. Dussan}{
  address={Department of Physics, Washington University, 
St. Louis, Missouri 63130, USA}
}
\author{A. Rios}{
  address={Department of Physics, Faculty of Engineering and Physical Sciences, 
University of Surrey, 
Guildford, Surrey GU2 7XH, United Kingdom}
}
\author{A. Polls}{
  address={Departament d'Estructura i Constituents de la Mat\`eria,
Universitat de Barcelona, E-08028 Barcelona, Spain}
}
\begin{abstract}
The influence of short-range correlations on the spectral distribution of neutrons is incorporated in the solution of the gap equation for the ${}^3P_2-{}^3F_2$ coupled channel in pure neutron matter.
This effect is studied for different realistic interactions including one based on chiral perturbation theory.
The gap in this channel vanishes at all relevant densities due to the treatment of these correlations.
We also consider the effect of long-range correlations by including polarization terms in addition to the bare interaction which allow the neutrons to exchange density and spin fluctuations governed by the strength of Landau parameters allowed to have reasonable values consistent with the available literature.
Preliminary results indicate that reasonable values of these parameters do not generate a gap in the ${}^3P_2-{}^3F_2$ coupled channel either for all three realistic interactions although the pairing interaction becomes slightly more attractive.
\end{abstract}

\maketitle


\section{Introduction}

The cooling scenario of a neutron star depends sensitively on the possible superfluidity of neutrons and superconductivity of protons in the interior of the system~\cite{Page2004}.
Pulsar glitches are still considered prime evidence of superfluid properties on account of their long relaxation times~\cite{Migdal59,Chamel08}.
Recent observations of a young neutron star in Cassiopeia A~\cite{Ho09} and its rapid cooling~\cite{Shternin11} have rekindled interest in high-density neutron superfluidity.
Several scenarios with apparently different mechanisms appear to be able to explain the observed cooling observation~\cite{Page11,Blaschke13} but both require a contribution of neutron pairing with ${}^3P_2-^3F_2$ quantum numbers.
Longer observation of this system may shed further light on this issue.
Meanwhile, it appears worthwhile to consider the present status of high-density neutron pairing.
At low density there is universal agreement that neutron pairs with ${}^1S_0$ quantum numbers will be superfluid~\cite{Dean03}.
These quantum numbers are obvious candidates for a pairing solution because the corresponding phase shift of the isovector partial wave for ${}^1S_0$ nucleons is attractive.
At higher energy, the $P$-wave phase shift of the coupled ${}^3P_2-^3F_2$ channel provides more attraction and this channel is therefore a candidate for neutron pairing at higher density.
At low densities all calculations based on realistic nucleon-nucleon (NN) interactions predict a homogeneous pairing phase characterized by ${}^1S_0$ neutron pairs as discussed in the review of Dean and Hjorth-Jensen~\cite{Dean03}.
At very low densities the superfluid gap is essentially determined by the corresponding phase shifts in this channel.
The BCS treatment of the gap equation with a spectrum of kinetic energies typically generates a maximum gap of about 3 MeV around a density corresponding to $k_F = 0.8\ \textrm{fm}^{-1}$ irrespective of the realistic interaction that is employed~\cite{Dean03}.
There is more variation in the gap obtained for the ${}^3P_2-^3F_2$ channel in corresponding calculations depending on the choice of the nucleon-nucleon interaction ranging from about 0.5 to 0.9 MeV for the maximum gap with a density range associated with Fermi momenta from 1.8 to 2.5 fm$^{-1}$.

Another example of an attractive interaction pertaining to a coupled channel occurs in symmetric nuclear matter for isoscalar pairs with ${}^3S_1$-${}^3D_1$ quantum numbers.
We will build on the experience with pairing in this coupled channel to motivate our treatment of neutron ${}^3P_2-^3F_2$ pairing.
We note that early calculations around normal nuclear matter density generated a sizable gap of around 10 MeV~\cite{ars:90,vgdpr:91,Baldo92,Takatsuka93,Baldo95} with ${}^3S_1$-${}^3D_1$ quantum numbers. 
Those calculations typically solve the BCS gap equation using a mean-field single-particle (sp) propagators with sp energies  \textit{e.g.} determined in a Brueckner-Hartree-Fock (BHF) calculation.
The empirical data of finite nuclei do not exhibit any indications for
such strong proton-neutron pairing correlations with corresponding gaps as large as 10 MeV. 
While finite size effects and the notion that the corresponding proton-neutron pairs have total spin $S = 1$ and therefore may be harder to accommodate in a system with closed shells, it is unlikely that a 10 MeV gap in nuclear matter at normal density would not have some observable consequences in finite systems. 
It is therefore plausible that the evaluation of the ${}^3S_1$-${}^3D_1$ gaps in infinite matter according to Refs.~\cite{ars:90,vgdpr:91,Baldo92,Takatsuka93,Baldo95} suffers from the lack of inclusion of relevant physics that can suppress such a large gap.

A candidate for such a mechanism is the notion that nucleons in nuclei are well documented to behave differently than expected on the basis of a pure independent particle model (IPM).
Such additional correlations beyond the Pauli principle of the IPM are clearly established in the analysis of the $(e,e'p)$ reaction of closed-shell nuclei.
For example the removal probability of valence protons in closed-shell nuclei is reduced to about 0.65 compared to an IPM value of 1~\cite{Lapikas93}.
A substantial contribution to this reduction, about 0.10 to 0.15, is a consequence of the depletion of the Fermi sea due to short-range correlations (SRC)~\cite{Dickhoff04}.
The complementary admixture of a modest amount of high-momentum components in the nuclear ground state confirming this interpretation has also been observed~\cite{Rohe2004}. 
Such correlations can be accurately and completely treated in infinite nuclear systems where the consequences of summing ladder diagrams can be included into the self-energy and off-shell propagation can be incorporated self-consistently~\cite{Dickhoff04}.
While such a procedure still encounters some numerical issues at zero temperature no such problems are encountered at sufficiently high temperature~\cite{libphd,frickphd,riosphd,Dickhoff2008}.
At finite temperature it is thus possible to fully account for the influence of SRC on the propagation properties of nucleons in symmetric nuclear and neutron matter.
At zero temperature ladder diagram summations can lead to pairing instabilities when the possibility of anomalous (pair)  propagation is not included.
Since the influence of pairing on the normal nucleon propagator is confined to energies around the Fermi energy that are of the order of the size of the gap, it is reasonable to assume that the normal self-energy will hardly be affected by pairing correlations.
Making this assumption and extrapolating the temperature dependence of the self-consistently calculated normal self-energy to lower temperatures including $T=0$, it was shown in Ref.~\cite{Muther2005} that the influence of SRC on  
${}^3S_1$-${}^3D_1$ pairing is quite dramatic when this effect is properly included in the gap equation. 
At normal density the gap then vanishes reconciling empirical information gathered from nuclei with many-body calculations of pairing in nuclear matter.
It is therefore natural to follow this procedure for the calculation of the influence of SRC on the pairing properties in neutron matter at high density for the ${}^3P_2-^3F_2$ coupled channel.
We note that such self-energy effects have also been considered by other authors~\cite{Bozek00,Baldo00,Baldo02,Bozek03} but it was shown that approximations with effective strength factors are not reliable~\cite{Muther2005}.
 
 Another important ingredient for the calculation of pairing correlations at high density in neutron matter is the modification of the pairing interaction when nucleons exchange the low-energy, possibly collective, excitations of the medium.
At small momentum transfer such excitations are determined by the Landau parameters that govern the density and spin excitations of the medium~\cite{Dickhoff2008}.
The first correction to be added to the bare NN interaction is given by the exchange of one particle-hole bubble between the neutrons.
This mechanism can be extended to exchanging an infinite order random-phase-approximation (RPA) bubble series.
Such contributions arise since the gap equation itself generates the contribution of ladder diagrams~\cite{Migdal1967,Dickhoff88,Dickhoff2008}.
The accurate calculation of Landau parameters depends on many ingredients some of which were discussed in Ref.~\cite{Dickhoff1987}.
We will also include the contribution of these so-called polarization terms that reflect long-range correlations (LRC) and follow the procedure outlined in Ref.~\cite{Cao2006} for neutron ${}^1S_0$ superfluidity and extend it to the case of the ${}^3P_2-^3F_2$ coupled channel.
 
We outline the relevant ingredients of the calculation in the next section and subsequently discuss the results.
 
\section{Formalism}

\subsection{Self-consistent ladder diagrams at finite $T$}
We start with a brief and schematic reminder of the self-consistent summation of ladder diagrams that was used for the construction of the ingredients of the pairing calculation~\cite{Rios2009a,Rios2014}.
The proper treatment of SRC and tensor correlations in the nuclear medium is accomplished by constructing the
in-medium Lippman-Schwinger equation, schematically represented by:
\begin{equation}
	\T = V + V \G_{II}^0 \T \, .
\label{eq:LSeq}
\end{equation}
The NN interaction is denoted by $V$ and the noninteracting but dressed two-particle propagator is given by
\begin{eqnarray}
\G_{II}^0(k,k'; \Omega) = 
\int \frac{\textrm{d} \omega}{2 \pi} \frac{\textrm{d} \omega'}{2 \pi} 
\frac{ \G^<(k,\omega)  \G^<(k',\omega') - \G^>(k,\omega)  \G^>(k',\omega')}{\Omega - \omega - \omega' +i\eta} \, .
\label{eq:g2}
\end{eqnarray}
The two components of the sp propagator, $\G^<$ and $\G^>$, are
related by a Kubo-Martin-Schwinger relation. They are also linked to the sp spectral function, $\A$:
\begin{eqnarray}
 \G^<(k,\omega) =  f(\omega) \A(k,\omega) \, ,\\
 \G^>(k,\omega) =  [1 - f(\omega)] \A(k,\omega) \,
 \label{eq:prop}
\end{eqnarray}
using the Fermi-Dirac distribution
\begin{equation} 
f(\omega) = \frac{1}{ 1 + e^{  ( \omega - \mu )/T} }.
\end{equation}
The temperature, $T$, is applied as an external condition. 
The chemical potential, $\mu$, is obtained from the normalization
to the density of the momentum distribution
\begin{equation}
n(k)= 
\int_{-\infty}^{+\infty} \frac{{\mathrm{d}}\omega}{2\pi}
f(\omega)
\A(k,\omega) .
\label{eq:momdis}
\end{equation}
For neutron matter we include a degeneracy factor $\nu = 2$ when calculating the density from the momentum distribution.
Self-consistency is imposed all the way through in our calculations. The 
sp propagators
that enter Eq.~(\ref{eq:g2}), for instance, are obtained from the $\T-$matrix itself. 
In the ladder approximation, 
this effective interaction defines the imaginary part of the self-energy:
\begin{eqnarray}
\textrm{Im}\ \Sigma (k,\omega) = 
 \int \!\!\! \frac{\textrm{d}^3 k'}{ (2 \pi)^3} \int \! \frac{\textrm{d}\omega'}{2 \pi}  
\left[ f(\omega') + b(\omega + \omega')  \right] 
  \left\langle \bm{k}  \bm{k}' \right| \textrm{Im} \T ( \omega + \omega' )\left|  \bm{k} \bm{k}' \right\rangle_A 
\A(k',\omega') \, .
\label{eq:imself}
\end{eqnarray}
which receives different contributions that arise from both the 
fermionic, $f(\omega)$, and bosonic
\begin{equation}
\label{eq:boson}
b(\Omega) = \frac{1}{ e^{  ( \Omega - 2\mu)/T}  -1 },
\end{equation}
phase-space factors. 
Antisymmetrized matrix elements of $\T$ are included denoted by the $A$ subscript in Eq.~(\ref{eq:imself}).

The dispersive contribution to the real part of the self-energy can be obtained from
a dispersion relation~\cite{Dickhoff2008}. 
In addition the energy-independent correlated Hartree-Fock contribution, 
\begin{eqnarray}
\Sigma^{HF}(k) = 
\int \!\!\! \frac{\textrm{d}^3 k'}{ (2 \pi)^3} \int \! \frac{\textrm{d}\omega'}{2 \pi}  
\left\langle \bm{k} \bm{k}' \right| V \left|  \bm{k} \bm{k}' \right\rangle_A 
 G^<(k',\omega') \, .
\label{eq:self_HF}
\end{eqnarray}
must be included as well. 
We consider up to $J=8$ partial waves in the calculation of this contribution, which is relevant
for in-medium quasi-particle shifts. 
From Dyson's equation we then obtain the sp spectral function:
\begin{equation}
\A(k,\omega) = \frac{-2 \textrm{Im} \Sigma(k,\omega) }
{ [ \omega - \frac{k^2}{2m} - \textrm{Re} \Sigma(k,\omega) ]^2+ [ \textrm{Im} \Sigma(k,\omega) ]^2} \, ,
\label{eq:asf}
\end{equation}
which closes the self-consistency loop that treats all particles in the same manner, 
providing appropriate feedback to the different ingredients of the calculations.

\subsection{Pairing treatment beyond BCS}

The proper treatment of pairing requires the introduction of anomalous propagators~\cite{Dickhoff2008} with corresponding anomalous self-energy terms in so-called Gorkov equations~\cite{Gorkov58,Migdal1967} that couple normal and abnormal (pair) propagators. 
The resulting expression for the anomalous self-energy $\Delta$ can be rewritten~\cite{Bozek99}
in a partial wave expansion
\begin{eqnarray}
\Delta^{JST}_{\ell}(p)  = \sum_{\ell'} \int_0^{\infty} 
\frac{{\mathrm{d}} k \,k^{2}}{(2\pi)^3}
\int_{-\infty}^{+\infty}\frac{{\mathrm{d}}\omega}{2\pi}
\int_{-\infty}^{+\infty}\frac{{\mathrm{d}}\omega'}{2\pi}
\left<p|V^{JST}_{\ell \ell'}|k\right>_A
\A(k,\omega) \A_s(k,\omega')
\frac{1-f(\omega)-f(\omega')}
{-\omega-\omega'}\Delta^{JST}_{\ell'}(k)\, ,
\label{eq:gappw}
\end{eqnarray}
where $\A_s$ is the spectral function that includes the effect of a nontrivial pairing solution to the gap equation.
If we ignore for a moment the difference between the spectral functions $\A$ and
$\A_s$, we realize that the equation for $\Delta$ corresponds to 
the homogeneous scattering equation for the $\T$-matrix in (\ref{eq:LSeq}) 
at energy $\Omega=0$ and center-of-mass momentum $P=0$. 
This means that a non-trivial solution of Eq.~(\ref{eq:gappw}) is
obtained if and
only if the scattering matrix $\T$ generates a pole at energy $\Omega=0$, which 
reflects a bound two-particle state. This is precisely the condition for the
pairing instability discussed in the introduction, demonstrating that this treatment of 
pairing correlations is compatible with the $\T$-matrix approximation in the
non-superfluid regime discussed in the previous subsection~\cite{Dickhoff88}. 


The spectral functions $\A(k,\omega)$ and $\A_s(k,\omega)$ in Eq.~(\ref{eq:gappw}) are usually approximated
by the corresponding mean-field and BCS approximation. 
The normal spectral function then reads
\begin{equation}
\A(k,\omega)=2\pi\delta(\omega + \mu -\varepsilon_k) =2\pi\delta(\omega - \chi_k)
\,,\label{eq:specqp}
\end{equation}
with the quasi-particle energy $\varepsilon_k$ for a mean-field particle with momentum $k$
and $\chi_k= \varepsilon_k - \mu$. Note, that for convenience we define the
energy variable relative to the chemical potential $\mu$.
The BCS
approximation for the spectral function yields
\begin{equation}
\A_s(k,\omega) = 2\pi\left(\frac{E_k+\chi_k}{2E_k}\delta(\omega-E_k) +
\frac{E_k-\chi_k}{2E_k}\delta(\omega+E_k)\right),\label{eq:specbcs}
\end{equation}
with the quasi-particle energy
\begin{equation}
E_k = \sqrt{\chi_k^2 + \Delta^2(k)}\,.\label{eq:eqp}
\end{equation} 
Inserting these approximations for the spectral function into
Eq.~(\ref{eq:gappw}) and taking the limit $T=0$ reduces to the usual
BCS gap equation
\begin{equation}
\Delta^{JST}_\ell(p) = - \sum_{\ell'} \int_0^{\infty} 
\frac{{\mathrm{d}} k\,k^{2}}{(2\pi)^3} \left<p|V^{JST}_{\ell \ell'}|k\right>_A \frac{1}{2E_k}
\Delta^{JST}_{\ell'}(k) \,.
\label{eq:gapbcs}
\end{equation}
It is therefore appropriate to consider Eq.~(\ref{eq:gappw}) as a generalization of the usual
gap equation. It accounts for the spreading of sp strength due to SRC and tensor correlations~\cite{Dewulf2003,Rios2009a} employing a generalization of the form
\begin{equation}
-\frac{1}{2E_k}\quad \to \quad  
\int_{-\infty}^{+\infty}\frac{{\mathrm{d}}\omega}{2\pi}
\int_{-\infty}^{+\infty}\frac{{\mathrm{d}}\omega^{\prime}}{2\pi} 
\A(k,\omega) \A_s(k,\omega^{\prime})
\frac{1-f(\omega)-f(\omega^{\prime})}
{-\omega-\omega^{\prime}}\,.\label{def:zk0}
\end{equation}
For practical calculations we have assumed that different partial-wave projections do not couple through the angular dependence of the interaction and the propagators.
We also employ the spherical version of the quasiparticle energy in the case of coupled channels facilitated by writing
\begin{equation}
\Delta^{JST}(k) = \sqrt{\left(\Delta^{JST}_{\ell=J-1}(k)\right)^2+\left(\Delta^{JST}_{\ell=J+1}(k)\right)^2} .
\label{eq:deltaJST}
\end{equation}

We will also report preliminary results for the inclusion of polarization terms in the pairing interaction.
We have followed the procedure proposed in Ref.~\cite{Cao2006} and generalized it to the case of the ${}^3P_2-^3F_2$ coupled channel.
Since these results are preliminary we refer for further details at this time to Ref.~\cite{Cao2006}.
We do note that the limitation to the use of Landau parameters is of some concern on account of the presence of the pion-exchange tensor interaction that can influence the spin mode and may exhibit quite different behavior as a function of momentum transfer~\cite{Dickhoff1982}.
Additional effects may be expected since density or spin modes should in principle depend on the excitation energy making such polarization contributions energy dependent.
These features naturally lead to a complex and energy-dependent gap function.

\section{Results and Discussion}

The essential calculation that provides the input for the solution of gap equation of Eq.~(\ref{eq:gappw}) is provided by a proper temperature extrapolation of the convolution of spectral functions generated by the self-consistent normal self-energy given by
\begin{equation}
\frac{1}{-2\widetilde\chi_k} = :
\int_{-\infty}^{+\infty}\frac{{\mathrm{d}}\omega}{2\pi}
\int_{-\infty}^{+\infty}\frac{{\mathrm{d}}\omega^{\prime}}{2\pi} 
\A(k,\omega) \A(k,\omega^{\prime})
\frac{1-f(\omega)-f(\omega^{\prime})}
{-\omega-\omega^{\prime}} \,\label{def:zk1}
\end{equation}
at a given temperature.
This convolution provides the generalization of  $\chi_k$ introduced in Eq.~(\ref{eq:specqp}) relevant for the mean-field limit and the conventional form of the BCS gap equation.
For the present work we have explored an expansion of both the real and imaginary part of the normal self-energy in even powers of the temperature.
This expansion is then constrained by relevant macroscopic thermodynamical quantities.
The zero-temperature limit can then be used to construct the relevant spectral functions that are needed for the calculation of Eq.~(\ref{def:zk1}).

\begin{figure}
  \includegraphics[height=.3\textheight]{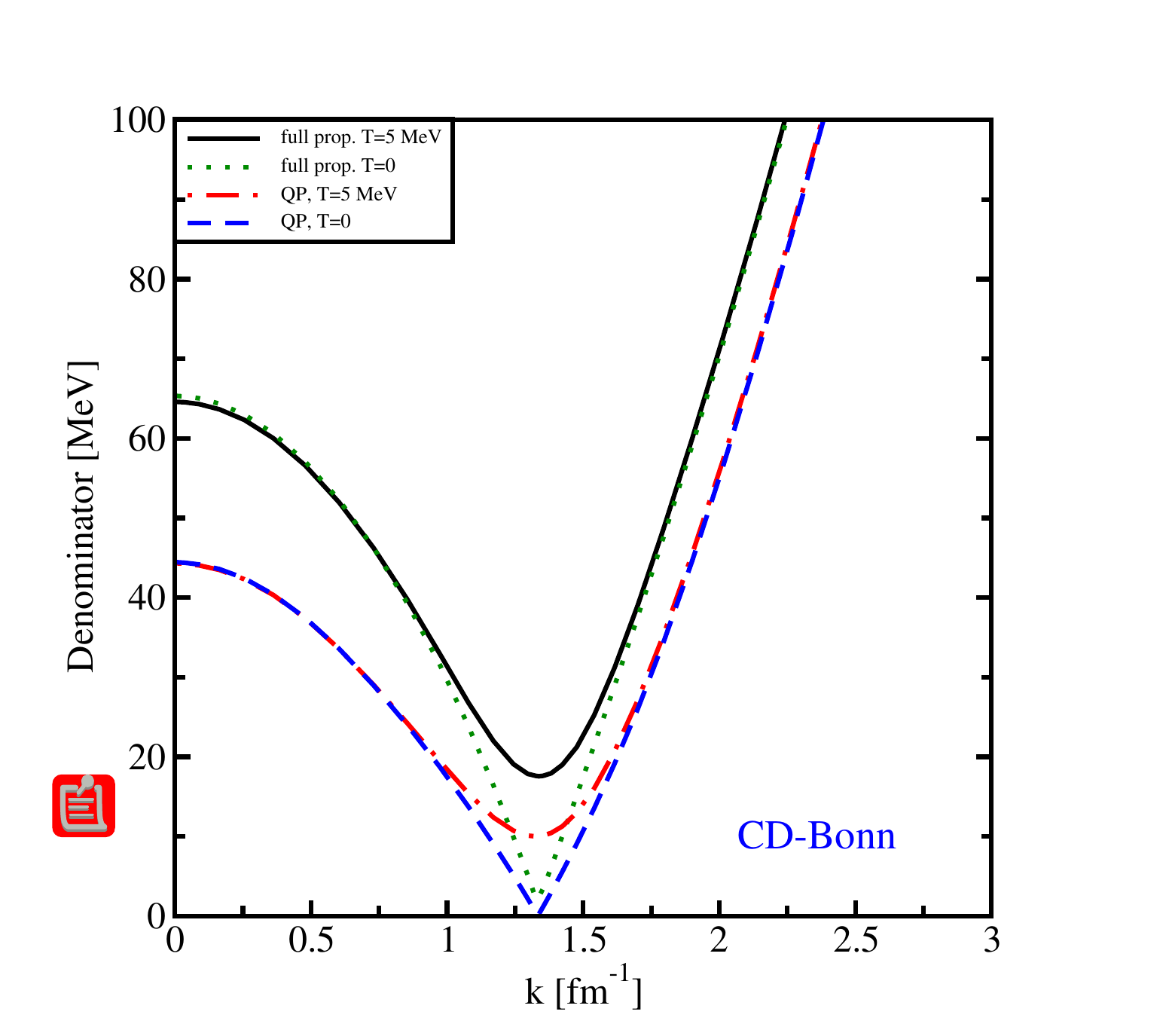}
  \caption{Result of the convolution of Eq.~(\ref{def:zk1}) at normal density in symmetric nuclear matter. }
\label{fig:chi}
\end{figure}

We display in Fig.~\ref{fig:chi} the result of such a procedure at normal density (0.16 fm$^{-3}$) for symmetric nuclear matter. 
The solid line represent the convolution directly calculated at $T = 5$ MeV.
Using the quasiparticle-energy approximation, representing only the location of the peak of the spectral functions, leads to the dash-dotted line at this temperature.
The result of the constrained extrapolation to $T=0$ of the self-energy and subsequent convolution is represented by the dotted line.
The corresponding quasiparticle result by the dashed curve.
The difference between the quasiparticle approximation and the full convolution clarifies the reduced tendency to generate a solution of the gap equation in the latter case.
It should also be noted that the momentum dependence is substantially different.  
The results displayed in Fig.~\ref{fig:chi} are essentially equivalent to Fig.~5 of Ref.~\cite{Muther2005}.
We emphasize that the self-consistent inclusion of SRC in the convolution of Eq.~(\ref{def:zk1}) reconciles the notion that proton-neutron pairing which is apparently not very relevant for finite nuclei, is correspondingly confined to densities below the one observed in the interior of nuclei when properly calculated in nuclear matter, \textit{i.e.} with the inclusion of SRC in the normal spectral functions.
Applying the same approach to higher densities in neutron matter is therefore a sensible strategy.

\begin{figure}[b]
  \includegraphics[height=.3\textheight]{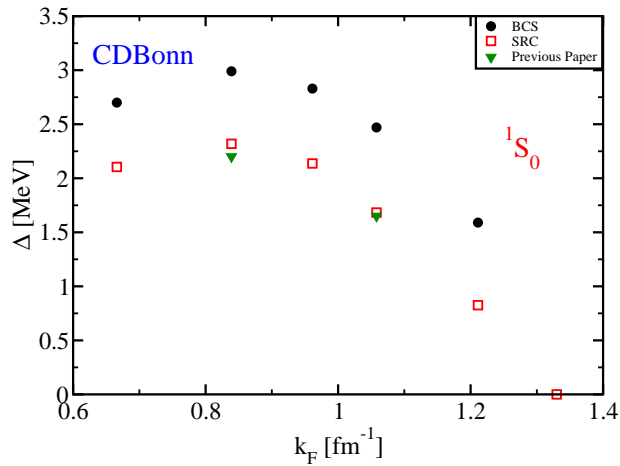}
  \caption{Gap for ${}^1S_0$ neutron pairing as a function of the Fermi momentum in pure neutron matter. For symbols see the text.}
\label{fig:gap1S0}
\end{figure}
We first show in Fig.~\ref{fig:gap1S0} the gap for neutron matter with ${}^1S_0$ quantum numbers as a function of the Fermi momentum calculated with the charge-dependent Bonn (CD-Bonn) interaction~\cite{Machleidt1995}.
The solid circles refer to the gap obtained from a standard BCS gap equation employing sp  energies that only contain kinetic energy terms.
The open squares are obtained by employing the convolution of Eq.~(\ref{def:zk1}) to obtain the generalized $\tilde{\chi}_k$ function.
The inverted triangles correspond to the results of Ref.~\cite{Muther2005} using the same interaction but a different extrapolation procedure to arrive at the $\tilde{\chi}_k$ function at $T=0$.
We conclude that these different implementations yield consistent results within about 0.1 MeV for this channel.
The main effect of the inclusion of SRC is to reduce the value of the gap by about 0.75 MeV and generate a gap closure at a corresponding lower density.
Similar results are obtained for the Idaho next-to-next-to-next-to-leading-order
(N3LO) chiral-perturbation-theory potential of Ref.~\cite{Entem2003} and the Argonne v18 (Av18) interaction of Ref.~\cite{Wiringa1995}.
We note that the inclusion of self-energy effects due to SRC removes sp strength from the vicinity of the Fermi energy and distributes it far and wide without compensating the loss of strength near the Fermi energy.
This feature has the inescapable consequence that a pairing solution is less likely and therefore the gap is suitably reduced or vanishes.
Other calculations that take such effects into account yield identical conclusions~\cite{Bozek00,Baldo00,Baldo02,Bozek03}.

\begin{figure}[tb]
  \includegraphics[height=.3\textheight]{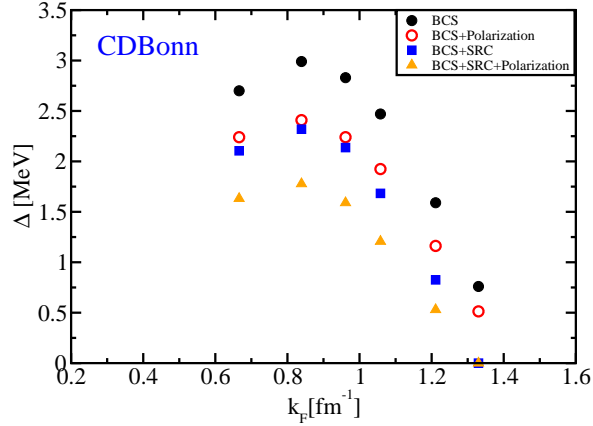}
  \caption{Gap for ${}^1S_0$ neutron pairing as a function of the Fermi momentum in pure neutron matter. For symbols see the text.}
\label{fig:gap1S0L}
\end{figure}
Calculations for the ${}^1S_0$ channel are readily available. 
For a list of references see for example Ref.~\cite{Gandolfi2009}.
In essentially all cases a reduction of the gap is obtained either due to the treatment of SRC or when LRC are included representing the modification of the pairing interaction due to the exchange of low-energy density and spin excitations.
We have followed the procedure and employ the Landau parameters of Ref.~\cite{Cao2006} to include such LRC in the pairing interaction.
The full circles of Fig.~\ref{fig:gap1S0L} denote again the BCS gap with kinetic energies.
Full squares include only the effect of SRC while open circles include only the effect of LRC.
We have compared our results due to the inclusion of LRC with those of Ref.~\cite{Cao2006} and have found that they are consistent.
The triangles include in the calculation of the gap both the effect of SRC and LRC.
The results are shown for the CD-Bonn interaction but the N3LO and Av18 interaction generate very similar results characterized by a maximum gap of about 1.7 MeV at a density corresponding to a Fermi momentum of a little more than 0.8 fm${}^{-1}$.

\begin{figure}[tb]
  \includegraphics[height=.3\textheight]{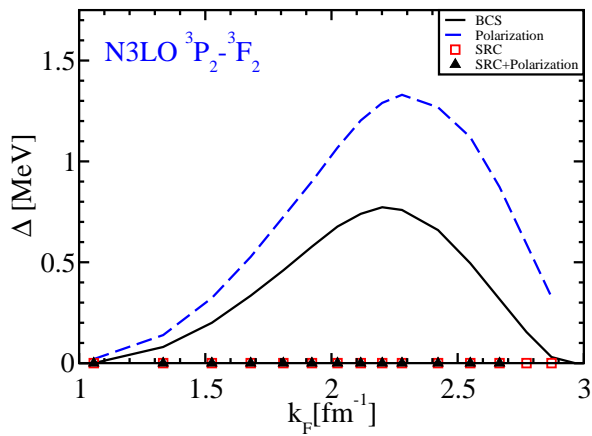}
  \caption{Gap for ${}^3P_3-^3F_2$ neutron pairing as a function of the Fermi momentum in pure neutron matter. For symbols see the text.}
\label{fig:gap3PF2}
\end{figure}
The influence of polarization contributions in the ${}^1S_0$ channel is known to reduce the pairing gap and is therefore considered a screening effect.
In the case of the ${}^3P_3-^3F_2$ channel, the spin recoupling from particle-hole to particle-particle leads to an interaction that enhances the pairing gap and can therefore be characterized as anti screening~\cite{Ding2014}.
The results including polarization contributions are at this stage of our work still preliminary so we will include details in our future publication~\cite{Ding2014}.
The solid line in Fig.~\ref{fig:gap3PF2} refers to the solution of the BCS gap equation with kinetic sp energies for the N3LO interaction. 
When only the SRC influence is added to the calculation no nontrivial solution of the gap equation is obtained which is identified by the open squares in the figure.
With the addition of polarization terms but without SRC effect the result is given by the dashed curve demonstrating the additional attraction that raised the maximum gap from the BCS value of about 0.75 MeV to 1.3 MeV at a slightly higher density.
The simultaneous inclusion of SRC and LRC in its preliminary form suggests that there is still no non-vanishing solution of the generalized gap equation as identified by the filled triangles in Fig.~\ref{fig:gap3PF2}.
Since these results are preliminary, it is premature to draw definite conclusions at this time.
Other NN interactions yield already different results for the gap when the standard BCS calculation with kinetic energies is performed~\cite{Baldo1995}.
In our case, we note that the CD-Bonn interaction yields a similar maximum gap of about 0.8 MeV at the BCS level but at a substantially higher density. 
The influence of LRC in this case is then larger but the effect or SRC is still such that no gap survives in this case either.
The BCS result for the Av18 interaction is also different and the LRC effect is smaller as well but again the inclusion of SRC yields no longer a nontrivial solution to the gap equation.

We note that a recent publication studied the ${}^3P_3-^3F_2$ gap equation with the inclusion of three-body forces that are transformed to a density dependent two-body force~\cite{Dong2013}.
In Ref.~\cite{Dong2013} the Bonn B interaction was employed for the NN interaction~\cite{Machleidt1987} which generates a very small maximum gap of about 0.25 MeV in the BCS case but using a Brueckner-Hartree-Fock sp spectrum.
Including SRC approximately by using the strength factor renormalization, a very small maximum gap of 0.04 MeV is obtained which is slightly increased when the effect of three-body forces is included.

We may first of all conclude that the inclusion of SRC has a dramatic but not unexpected effect on the solution of the gap equation in high-density neutron matter for the ${}^3P_3-^3F_2$ coupled channel.
This conclusion is valid for the hard Av18, the softer CD-Bonn, as well as the very soft N3LO interaction.
Considering our preliminary results for the inclusion of LRC suggests that SRC continue to dominate and no nontrivial solution of the gap equation may occur.
The future inclusion of three-body forces associated with the N3LO interaction appears possible~\cite{Carbone2013a,carbonephd} but may not change these conclusions based on the results of Ref.~\cite{Dong2013}.
If these conclusions hold, they have important implications for the cooling scenarios of neutron stars.
We look forward to continuing our study of the problem of neutron pairing at high densities.



\begin{theacknowledgments}
This work is partly supported by by the Consolider
Ingenio 2010 Programme CPAN CSD2007-00042, Grant
No. FIS2011-24154 from MICINN (Spain), and Grant No.
2009GR-1289 from Generalitat de Catalunya (Spain); by
STFC, through GrantsNo. ST/I005528/1 and No. ST/J000051; by NewCompstar, COST Action MP1304, 
and by the US National Science Foundation under Grants No.
PHY-0968941 and No. PHY-1304242.
\end{theacknowledgments}



\bibliographystyle{aipproc}   


\bibliography{biblio3PF2}


\end{document}